# 基于多分辨率 RPCA 的互联网流量矩阵结构分析

王哲　　胡凯　　尹宝林

(北京航空航天大学计算机学院 软件开发环境国家重点实验室 北京 100191)

**摘　要**　互联网流量矩阵在网络运营与管理中发挥着重要作用，流量矩阵结构分析分解高维流量数据的各种流量成分，具有重要应用价值。本文基于 Robust Principal Component Analysis（RPCA）理论，并结合小波多分辨率分析，提出一种新的流量矩阵结构分析手段——多分辨率 RPCA。首先，建立多分辨率流量矩阵分解模型，利用小波系数刻画确定性流量的光滑特性。其次，改进 Stable Principal Component Pursuit（SPCP），使用多分辨率约束改造新的流量矩阵分解方法 SPCP-MRC，并设计数值算法。针对数值算法的一个子问题，给出并严格证明了解析解。最后，模拟多组包含各种网络异常与不同幅度噪声的流量矩阵用于实验评价，比较 SPCP-MRC 与多种流量分解方法，验证此流量分解结果具有更高的准确性与合理性。

**关键词**　流量矩阵；RPCA；多分辨率分析；数值算法；解析解；模拟实验
**中图法分类号**　TP393

## Internet Traffic Matrix Structural Analysis Based on Multi-Resolution RPCA

WANG Zhe　　HU Kai　　YIN Baolin

(State Key Laboratory of Software Development Environment, School of Computer Science and Technology, Beihang University, Beijing 100191, China)

**Abstract**　The Internet traffic matrix plays a significant roll in network operation and management, therefore, the structural analysis of traffic matrix, which decomposes different traffic components of this high-dimensional traffic dataset, is quite valuable to some network applications. In this study, based on the Robust Principal Component Analysis (RPCA) theory, a novel traffic matrix structural analysis approach named Multi-Resolution RPCA is created, which utilizes the wavelet multi-resolution analysis. Firstly, we build the Multi-Resolution Traffic Matrix Decomposition Model (MR-TMDM), which characterizes the smoothness of the deterministic traffic by its wavelet coefficients. Secondly, based on this model, we improve the Stable Principal Component Pursuit (SPCP), propose a new traffic matrix decomposition method named SPCP-MRC with Multi-Resolution Constraints, and design its numerical algorithm. Specifically, we give and prove the closed-form solution to a sub-problem in the algorithm. Lastly, we evaluate different traffic decomposition methods by multiple groups of simulated traffic matrices containing different kinds of anomalies and distinct noise levels. It is demonstrated that SPCP-MRC, compared with other methods, achieves more accurate and more reasonable traffic decompositions.

**Key words**　Traffic Matrix; RPCA; Multi-Resolution Analysis; Numerical Algorithm; Closed-Form Solution; Simulation Experiment

## 1　引言

　　互联网流量数据中蕴含丰富的网络行为信息，是网络运营与管理依赖的重要数据源，因此互联网流量分析是重要的研究课题。已有的流量分析研究多数针对单一链路流量数据[1]，难以感知全网络流量变化态势；近年来，互联网流量测量技术的进步，特别是支持 netflow 的骨干网路由器广泛部署，有效的支持分布式多点采集全网络级别的流量数据。全网络级流量记录一段时间内网络多处的流量属性，可以视为一个高维数据集，其中每个测量周期收集的全部流量属性视为一个高维数据点在各维度上的坐标值。

流量矩阵是常用的全网络级流量数据，记录被测量网络所有源-目的（Origin-Destination，OD）节点对间传输的流量值[1]，被广泛应用于流量工程[2]、全网络异常检测[3]等应用问题中。相比单一链路流量，流量矩阵包含网络多个位置的流量信息，可以更有效指导网络运营与管理；但由于流量矩阵数据具有高维特性，对其进行认知与分析通常也更为困难。

互联网是典型的开放系统[4]，互联网流量由大量用户多种通信行为共同产生，因此可以视为多种流量成分的叠加，其中包括代表"日模式"等周期性行为的正常流量、代表网络攻击、突发访问、故障等行为的异常流量、代表用户小幅度随机波动行为的噪声流量等。各种流量成分在网络管理中发挥着不同的作用，但通常测量到的是总流量数据。从互联网总流量中准确分解各种流量成分，是流量分析研究中的一个重要问题，文献[5]中称为流量结构分析。对一维流量时间序列的分解有不少成熟的数学工具，如 Fourier 分析、小波分析[6]等；然而这些方法大多难以推广到高维流量矩阵分析。

某些高维数据集可以通过恰当数学变换提取关键特征，并利用低维数据集近似，这些变换称为数据降维法。PCA[5]、NMF[7]等降维法已被应用于高维流量数据分解，研究发现正常流量矩阵具有低秩特性，且当异常流量幅度较小时，可通过总流量投影到几个最大奇异值主成分方向近似。然而，后续研究[8][9]说明当异常流量幅度较大时，PCA 的鲁棒性较差，不能有效分离正常流量和异常流量。近年来，RPCA（Robust Principal Component Analysis）理论[10]研究取得重要突破，并在计算机科学多个领域得到成功应用。RPCA 理论是对经典 PCA 的改进，对大幅度稀疏干扰具有很高的鲁棒性。最近，RPCA 也被用于分析互联网高维流量数据[11][12][13][14]，能够更准确分解低秩的正常流量，并且分离稀疏的异常流量。

尽管如此，基于 RPCA 的高维流量数据结构分析也存在一些不足，表现在流量分解结果的时间特性上：(1)正常流量刻画互联网流量的长期性趋势，故随时间缓慢光滑变化，但基于 RPCA 的分解结果不具备光滑性；(2)噪声流量的随机波动应具有平稳特性，但分解结果中包含低频的周期性模式[13]。这些不足说明流量分解的准确性有待提高，其根源在于 RPCA 的问题模型不关注数据样本的时间特性。以合理的流量矩阵模型为出发点，将流量时间特性与 RPCA 理论融合设计流量矩阵结构分析方法，是研究的核心思路。本文首先对流量矩阵建立更精细的数学模型，合理刻画流量成分的时间特性，然后基于模型假设对 RPCA 优化问题进行修改，并设计新的流量矩阵分解数值算法。另外，已有流量分析研究多使用真实流量矩阵，但是从真实流量中无法获得各流量成分的 Ground-Truth 信息，难以量化的评价和比较不同流量分解方法的准确性。本文研究依赖模拟流量矩阵对多种流量分解方法进行大量实验和定量评价。

本文的贡献主要包括：(1)利用小波多分辨率分析工具，提出一种多分辨率流量矩阵分解模型，该模型是对文献[12]中流量矩阵分解模型的重要改进，利用小波分解系数合理刻画正常流量的光滑特性；(2)基于多分辨率流量矩阵分解模型，研究多分辨率约束的 RPCA 问题，提出新的流量矩阵分解方法 SPCP-MRC 及高效数值算法，特别是针对数值算法的某个子问题，给出并证明了解析解；(3)模拟了包含多种典型异常流量和不同幅度噪声流量的流量矩阵用于实验，将 SPCP-MRC 方法与其他流量分解方法进行不同角度比较，验证其具有更高的准确性。

---

1  一组 OD 对间传递的流量通常称为一个 OD 流。

## 2 相关工作

### 2.1 互联网流量矩阵模型

设 $X \in \mathbb{R}^{T \times P}$ 表示一个流量矩阵数据，其中 $P$ 为网络中 OD 流数目，$T$ 为时间周期数目。每个列向量 $X_p \in \mathbb{R}^T$ $(1 \le p \le P)$ 是一个 OD 流的时间序列；每个行向量 $X(t,:) \in \mathbb{R}^P$ $(1 \le t \le T)$ 是一个时间周期内网络中各 OD 流的流量值，也被看作一个高维数据点。流量矩阵数据兼具时间、空间特性：时间特性描述每个 OD 流随时间变化，空间特性描述不同 OD 流分布与相关性。文献[15]提出流量矩阵空间分布的重力模型，但未考虑 OD 流的时间特性，也没有刻画总流量的不同组成成分。文献[16]讨论单个 OD 流的组成模型，假设其包含周期性的趋势成分和小幅度随机波动的噪声成分，但是没有考虑可能存在的异常成分。文献[5]利用 PCA 方法分析流量矩阵，指出全部 OD 正常流量可视为低秩矩阵，并解释为正常流量相关度很高。文献[17, 18]分析和模拟了多种异常流量，研究表明异常流量在时间和空间都具有偶发性。文献[12]综合已有研究结果，提出流量矩阵分解模型（Traffic Matrix Decomposition Model，TMDM），将总流量矩阵 $X$ 视为三个流量子矩阵之和：

$$X = A + E + N, \tag{1}$$

其中 $A$ 表示低秩的确定性流量矩阵，$E$ 表示稀疏的异常流量矩阵，$N$ 表示随机噪声流量矩阵，经过噪声标准化，TMDM 模型与广义 RPCA 问题[19]模型假设等价；研究中还通过对真实流量矩阵的分解与讨论，说明 TMDM 模型是一种简单且合理的流量矩阵数学模型。TMDM 模型仍可以进一步精细化，例如确定性流量作为总流量趋势成分，随时间缓慢光滑变化，但这个特性在模型假设中未被刻画。

### 2.2 RPCA 理论及应用

标准 RPCA 问题用自然语言描述为[10]：已知数据矩阵 $X \in \mathbb{R}^{m \times n}$ 是两个未知矩阵 $A_0, E_0$ 之和，其中 $A_0$ 是低秩矩阵，$E_0$ 是稀疏矩阵且非零元素绝对值任意大，由 $X$ 分解 $A_0, E_0$。该问题数学上对应于优化问题：

$$\min_{A, E \in \mathbb{R}^{T \times P}} \text{rank}(A) + \gamma \|E\|_0 \\ s.t. \quad X = A + E. \tag{2}$$

其中，$\text{rank}(\cdot)$ 表示矩阵的秩；$\|\cdot\|_0$ 表示矩阵非零元素数目，称为 $l_0$ 范数；常数 $\gamma > 0$ 平衡两个目标函数。问题(2)是一个非凸组合优化，没有有效的求解方法。受压缩感知[20]与低秩矩阵填充[21]研究启发，Chandrasekaran等[22]和Candes等[10]独立提出近似问题(2)的等式约束凸优化问题：

$$\min_{A, E \in \mathbb{R}^{T \times P}} \|A\|_* + \lambda \|E\|_1 \\ s.t. \quad X = A + E. \tag{3}$$

称为主成分追踪（Principal Component Pursuit，PCP）。问题(3)中，$\|\cdot\|_*$ 表示矩阵所有奇异值之和，称为原子范数；$\|\cdot\|_1$ 表示矩阵元素绝对值之和，称为 $l_1$ 范数；$\lambda > 0$ 为加权参数。研究证明[10, 22]，在较弱假设条件下，PCP能够以相当大概率准确分解低秩矩阵与稀疏矩阵。

很多实际问题中，满足"低秩+稀疏"假设的矩阵常混合小幅度稠密噪声，例如常见的高斯噪声。噪

声情况下稳定的矩阵分解，即广义 RPCA 问题描述为：已知数据矩阵 $X \in \mathbb{R}^{m \times n}$ 是三个未知矩阵 $A_0, E_0, N_0$ 之和，其中 $A_0$ 是低秩矩阵，$E_0$ 是稀疏矩阵且非零元素绝对值任意大，$N_0$ 是独立同分布随机噪声矩阵且 Frobenius 范数 $\|N\|_F \leq \delta$，由 $X$ 分解 $A_0, E_0, N_0$。Zhou 等[19]推广 PCP 方法，提出求解广义 RPCA 问题的稳定主成分追踪（Stable Principal Component Pursuit，SPCP）方法，考虑不等式约束凸优化问题：

$$\min_{A, E \in \mathbb{R}^{T \times P}} \|A\|_* + \lambda \|E\|_1 \tag{4}$$
$$s.t. \quad \|X - A - E\|_F \leq \delta.$$

并证明在与 PCP 方法类似假设下，SPCP 以相当大概率对广义 RPCA 问题的解给出稳定估计。

    RPCA 理论的应用很广，起初多针对图像与视频数据，如视频背景建模、人脸阴影消除[10]、多幅图片鲁棒对齐[23]、图像低秩纹理提取[24]等，此后在其它领域数据处理中亦得到良好应用，如文本关键词提取[25]、音乐去噪[26]等。最近，RPCA 及其改进方法也被应用于互联网高维流量数据分析[11-14]，Abdelkefi 等[11]和 Bandara 等[14]基于严格的"低秩+稀疏"模型假设，使用 PCP 分解高维流量数据；Wang 等[12]指出总流量中通常还包含噪声成分，使用 SPCP 给出合理流量分解；针对 SPCP 分解噪声流量中包含低频周期性模式的不足，使用频域正则方法进行了优化[13]。但是，这些研究都是 RPCA 的直接应用和经验改进，未对高维流量数据的时间特性进行建模。

## 3 多分辨率流量矩阵分解模型

    本节以 TMDM 模型[12]为基础，使用小波多分辨率分析工具，建立更精细的流量矩阵模型，称多分辨率流量矩阵分解模型（Multi-Resolution Traffic Matrix Decomposition Model，MR-TMDM），下面分别对三个流量子矩阵建模。

### 3.1 确定性流量模型

    首先，依 TMDM 模型的假设，确定性流量矩阵 $A \in \mathbb{R}^{T \times P}$ 是一个低秩矩阵。其次，$A$ 中每个列向量 $A_p \in \mathbb{R}^T$ $(1 \leq p \leq P)$ 记录 OD 流 $X_p \in \mathbb{R}^T$ 的确定性趋势成分，随时间变化缓慢且具有光滑性。文献[6]中，流量趋势通过对总流量进行多层小波分解，保留低分辨率尺度系数，再进行小波逆变换获得。受此工作启发，本文利用小波分解系数值分布，对确定性流量时间序列的光滑性建模。对于给定的一对尺度函数及小波母函数 $\varphi_0(t)$ 和 $\psi_0(t)$，$A_p \in \mathbb{R}^T$ 的 $J$ 层小波分解 $(0 < J \leq \lceil \log T \rceil)$ 记为[2]：

$$A_p(t) = \sum_k a_p(J, k)\varphi_{J,k}(t) + \sum_{j=1}^{J}\sum_k d_p(j, k)\psi_{j,k}(t), \tag{5}$$

其中

$$\begin{cases} \varphi_{J,k}(t) = 2^{-J/2}\varphi_0(2^{-J}t - k) \\ \psi_{j,k}(t) = 2^{-j/2}\psi_0(2^{-j}t - k) \quad 1 \leq j \leq J \end{cases}, \tag{6}$$

---

2 严格来说，小波分解是针对连续信号 $A_p(t) = \sum_k A_p(k)\text{sinc}(t-k)$ 进行的，离散信号 $A_p$ 可看作 $A_p(t)$ 的采样。

且构成一组正规正交基，$\{a_p(J,k)=\langle A_p,\varphi_{J,k}\rangle\}_k$ 与 $\{d_p(j,k)=\langle A_p,\psi_{j,k}\rangle\}_k$ 分别为 $A_p$ 第 $J$ 层尺度系数与第 $j$ 层小波系数 $(1\le j\le J)$。假设存在一对尺度函数与小波母函数 $\{\phi_0,\psi_0\}$ 及正整数 $0<q\le\lceil\log T\rceil$，每个列向量 $A_p$ 前 $q$ 层小波系数 $\bigcup_{j=1}^{q}\{d_p(j,k)\}_k$ 均为零。记 $V_q$ 是第 $q$ 层尺度函数 $\{\phi_{q,k}(t)\}_k$ 张成的子空间，则 $A_p$ 前 $q$ 层小波系数均为零也等价于 $A_p\in V_q$。

### 3.2 异常流量模型

不同种类的异常流量虽然具有各自的分布特点，但也具有一个共性：异常流量在时间和空间上都具有偶发性。因此，假设异常流量矩阵是一个稀疏矩阵。对于每种特定的异常流量，则根据该异常的分布特点，建立合理的数学模型，本文 6.1 节将对几种典型的异常流量进行建模。

### 3.3 噪声流量模型

总流量中除去确定性流量和异常流量，剩余部分称为噪声流量。噪声流量的幅度一般比较小，且每个噪声流量时间序列通常可以视为零均值的平稳随机过程，本文研究考虑一种最简单的情况：假设 $N_p$ 是零均值，方差为 $\sigma_p^2>0$ 的高斯白噪声 $(1\le p\le P)$。

综上讨论，满足 MR-TMDM 模型的流量矩阵分解，数学上描述为以下多分辨率 RPCA 问题：

**问题 1**（多分辨率 RPCA）：假设流量矩阵 $X\in\mathbb{R}^{T\times P}$ 是三个未知流量子矩阵之和，其中确定性流量矩阵 $A$ 是低秩矩阵，且存在一对尺度函数与小波函数 $\{\varphi_0,\psi_0\}$ 及正整数 $0<q\le\lceil\log T\rceil$，每个列向量 $A_p$ 前 $q$ 层小波系数均为零；异常流量 $E$ 是稀疏矩阵且非零元素绝对值任意大；噪声流量矩阵 $N$ 每个列向量 $N_p$ 是均值为零、方差为 $\sigma_p^2>0$ 的高斯白噪声；由 $X$ 分解 $A,E,N$。

## 4 多分辨率约束的 SPCP 方法

与广义 RPCA 问题相比，多分辨率 RPCA 问题对确定性流量矩阵 $A$ 附加了新的假设，因此是广义 RPCA 问题的一个特殊情况。尽管 SPCP 方法可以直接应用于多分辨率 RPCA，本文研究希望充分结合模型假设，设计更为有效的流量矩阵分解方法。针对确定性流量的低秩性与异常流量的稀疏性，新方法沿用 SPCP 方法的目标函数 $\|A\|_*+\lambda\|E\|_1$；针对新的假设，对确定性流量增加一个约束函数，另外对噪声流量使用新的约束函数代替 SPCP 中 Frobenius 范数约束。这两个约束函数作用于流量时间序列的小波分解系数，利用信号的多分辨率性质，故称新的分解方法为多分辨率约束的 SPCP（Stable Principal Component Pursuit with Multi-Resolution Constraints，SPCP-MRC）。下面分别讨论确定性流量和噪声流量的约束函数。

### 4.1 确定性流量的多分辨率约束

根据确定性流量的模型假设，直接推出多分辨率约束函数。设 $V_q$ 是 $\{\phi_{q,k}(t)\}_k$ 张成的子空间，对应于低分辨率。定义

$$V_q^{(P)} = \underbrace{V_q \times \cdots \times V_q}_{P} \tag{7}$$

为 $V_q$ 的 $P$ 次乘积空间。确定性流量矩阵 $A$ 所有列向量 $A_p$ ($1 \le p \le P$) 前 $q$ 层小波系数均为零，由 $V_q^{(P)}$ 含义该假设条件等价于 $A \in V_q^{(P)}$。设 $V_q^{(P)}$ 的指标函数为 $I_{V_q^{(P)}}(X)$：

$$I_{V_q^{(P)}}(X) = \begin{cases} 0 & X \in V_q^{(P)} \\ +\infty & X \notin V_q^{(P)} \end{cases} \qquad \forall X \in \mathbb{R}^{T \times P},$$

则确定性流量矩阵的多分辨率约束定义为等式 $I_{V_q^{(P)}}(A) = 0$。

## 4.2 噪声流量的多分辨率约束

根据噪声流量时间序列小波分解系数的统计特性，设计多分辨率约束函数。考虑 OD 流 $X_p$ ($1 \le p \le P$) 噪声分量 $N_p \in \mathbb{R}^T$ 的 $J$ 层 ($1 \le J \le \lceil \log(T) \rceil$) 离散小波变换：

$$W_J N_p = \left[ \left\{ a_p^N(J,k) \right\}_{k=1}^{L_J} \left\{ d_p^N(J,k) \right\}_{k=1}^{L_J} \cdots \left\{ d_p^N(1,k) \right\}_{k=1}^{L_1} \right]^T, \tag{8}$$

其中 $L_j$ 为第 $j$ 层小波系数长度 ($1 \le j \le J$)，且 $L_J$ 亦为第 $J$ 层尺度系数长度。由模型假设，$N_p$ 是均值为零，方差为 $\sigma_p^2 > 0$ 的高斯白噪声。根据文献[27, 28]结果，各层小波系数 $\{d_p(j,k)\}_k$ ($1 \le j \le J$) 及尺度系数 $\{a_p(J,k)\}_k$ 分别近似于零均值高斯白噪声，若 $\{\varphi_0, \psi_0\}$ 是具有较高阶消失矩（vanishing moment）的正交小波族，上述小波系数与尺度系数方差满足：

$$\begin{cases} \mathrm{var}\left(d_p^N(j,\cdot)\right) \approx \sigma_p^2, 1 \le j \le J \\ \mathrm{var}\left(a_p^N(J,\cdot)\right) \approx \sigma_p^2. \end{cases} \tag{9}$$

定义 $N_p$ 对 $J$ 层离散小波变换的加权矩阵 $K_p^J$：

$$K_p^J = \begin{pmatrix} \Xi_p^J & & & \\ & \Pi_p^J & & \\ & & \ddots & \\ & & & \Pi_p^1 \end{pmatrix}, \tag{10}$$

其中 $\Xi_p^J = \frac{1}{\sigma_p} I_{L_J \times L_J}$；$\Pi_p^j = \frac{1}{\sigma_p} I_{L_j \times L_j}$，$j = 1, ..., J$。则 $M_p = K_p^J W_J N_p$ 可近似为零均值、单位方差的高斯白噪声。基于 $l_\infty$ 范数定义噪声流量矩阵的多分辨率约束函数：

$$\left\| K_p^J W_J N_p \right\|_\infty \le \delta, \quad 1 \le p \le P. \tag{11}$$

其中常数 $\delta > 0$ 控制约束强度，确保 $M_p$ 中每个元素取值在区间 $[-\delta, \delta]$ 外概率足够小，例如可以选取 $\delta = 1.96(2.33)$，对应于高斯分布的 95(99)% 分位点。

使用多分辨率约束(11)代替SPCP方法(4)中噪声流量的Frobenius范数约束，主要思想在于：总流量的主体是确定性流量，因此能量主要集中在低分辨率上，基于Frobenius范数约束估计噪声流量的能量分布与总

流量正相关，但由(9)式可知噪声流量在不同分辨率能量均匀分布；多分辨率约束能够显式的对噪声流量在不同分辨率能量分布进行控制，过滤可能混入的低分辨率确定性流量。

根据上述讨论，构造噪声流量多分辨率约束，需要估计每个 OD 流中噪声流量的方差参数 $\sigma_p$。采用文献[29]中方法，设 $\{d_p(1,k)\}_k$ 为总流量时间序列 $X_p$ 的第 1 层小波系数（注意它与 $\{d_p^N(1,k)\}_k$ 不同），基于中位数绝对离差（Median Absolute Deviation，MAD）给出 $\sigma_p$ 的估计公式：

$$\hat{\sigma}_p = \frac{1}{0.6745} \mathrm{MAD}\{d_p(1,k)\},\tag{12}$$

其中

$$\mathrm{MAD}\{d_p(1,k)\} = \mathrm{median}\{\left|d_p(1,k) - \mathrm{median}\{d_p(1,s)\}\right|\}.\tag{13}$$

### 4.3 多分辨率约束的 SPCP

由4.1节和4.2节介绍的约束函数，多分辨率约束的SPCP（SPCP-MRC）考虑以下优化问题：

$$\min_{A,E,N \in \mathbb{R}^{T \times P}} \| A \|_* + \lambda \| E \|_1$$

$$s.t. \begin{cases} A + E + N = X \\ I_{V_q^{(P)}}(A) = 0 \\ I_{\mathrm{Box}(\delta)}(N) = 0 \end{cases}.\tag{14}$$

其中 $\mathrm{Box}(\delta) = \{N \in \mathbb{R}^{T \times P} : \| K_p^{q-1} W_p^{q-1} N_p \|_\infty \le \delta, 1 \le p \le P\}$ 表示约束函数(11)的可行集，容易验证 $\mathrm{Box}(\delta)$ 是 $\mathbb{R}^{T \times P}$ 的凸子集；$\lambda > 0$ 平衡两个目标函数，依文献[30]方法取 $\lambda = 1/\sqrt{\max(T,P)}$。SPCP-MRC与SPCP区别有两点：一是对确定性流量增加约束条件 $I_{V_q^{(P)}}(A) = 0$；二是对噪声流量使用新的多分辨率约束。

## 5 数值算法

SPCP-MRC 问题是一个约束凸优化，内点法等二阶数值算法的计算开销较大。针对 SPCP 问题的 APG（Accelerate Proximal Gradient）算法[31]和 ALM（Augmented Lagrange Multiplier）算法[32]具有较高收敛速度，它们都是一阶数值算法，迭代求解原问题的无约束松弛问题。本节给出 SPCP-MRC 问题的 APG 算法。

考虑优化问题(14)的无约束松弛：

$$\min_{A,E,N \in \mathbb{R}^{T \times P}} F(A,E,N) = \mu g(A,E,N) + f(A,E,N)\tag{15}$$

其中

$$\begin{cases} g(A,E,N) = \| A \|_* + \lambda \| E \|_1 + I_{V_q^{(P)}}(A) + I_{\mathrm{Box}(\delta)}(N) \\ f(A,E,N) = \frac{1}{2} \| A + E + N - X \|_F^2 \end{cases}.\tag{16}$$

$f(A,E,N)$ 是凸函数且可微，是对等式约束 $A+E+N=X$ 的惩罚项；$g(A,E,N)$ 是 4 不可微凸函数的线性组合。$\mu>0$ 为松弛参数，当 $\mu\to 0$，优化问题(15)的解逼近SPCP-MRC问题(14)的解。设 $\nabla f$ 为 $f$ 的梯度，容易算出

$$\nabla f(A,E,N)=\begin{bmatrix} A+E+N-X \\ A+E+N-X \\ A+E+N-X \end{bmatrix},$$

且 $\nabla f$ 的 Lipschitz 常数为 $L=3$。

APG 算法每步迭代极小化函数 $F(A,E,N)$ 在特殊点 $(Y_k^A,Y_k^E,Y_k^N)\in\mathbb{R}^{T\times P}\times\mathbb{R}^{T\times P}\times\mathbb{R}^{T\times P}$ 的二阶近似函数 $Q(A,E,N,Y_k^A,Y_k^E,Y_k^N)$：

$$\begin{aligned} Q(A,E,N,Y_k^A,Y_k^E,Y_k^N)=\mu g(A,E,N)+f(Y_k^A,Y_k^E,Y_k^N)+\frac{L}{2}\left\|(A,E,N)-(Y_k^A,Y_k^E,Y_k^N)\right\|_F^2+ \\ \left\langle\nabla f(Y_k^A,Y_k^E,Y_k^N),(A,E,N)-(Y_k^A,Y_k^E,Y_k^N)\right\rangle \end{aligned} \tag{17}$$

$(Y_k^A,Y_k^E,Y_k^N)$ 选取使用 Nesterov 技巧，随迭代步数 $k$ 不断更新（详见算法 1）。容易推导出

$$\begin{aligned} &\min_{A,E,N\in\mathbb{R}^{T\times P}}Q(A,E,N,Y_k^A,Y_k^E,Y_k^N) \\ &=\min_{A\in\mathbb{R}^{T\times P}}\left\{I_{V_q^{(P)}}(A)+\mu\|A\|_*+\frac{L}{2}\left\|A-G_k^A\right\|_F^2\right\}+\min_{E\in\mathbb{R}^{T\times P}}\left\{\mu\lambda\|E\|_1+\frac{L}{2}\left\|E-G_k^E\right\|_F^2\right\}+ \\ &\quad\min_{N\in\mathbb{R}^{T\times P}}\left\{\mu I_{\text{Box}(\delta)}(N)+\frac{L}{2}\left\|N-G_k^N\right\|_F^2\right\}+C \end{aligned} \tag{18}$$

其中

$$\begin{cases} G_k^A=Y_k^A-\dfrac{L}{2}(Y_k^A+Y_k^E+Y_k^N-N) \\ G_k^E=Y_k^E-\dfrac{L}{2}(Y_k^A+Y_k^E+Y_k^N-N), \\ G_k^N=Y_k^N-\dfrac{L}{2}(Y_k^A+Y_k^E+Y_k^N-N) \end{cases}$$

且 $C$ 为常数，故原始极小化问题分裂为三个独立的优化子问题。根据文献[33]，这三个子问题分别等价于凸函数 $\dfrac{1}{L}I_{V_q^{(P)}}(A)+\dfrac{\mu}{L}\|A\|_*$，$\dfrac{\mu\lambda}{L}\|E\|_1$ 和 $\dfrac{\mu}{L}I_{\text{Box}(\delta)}(N)$ 对应的 Proximal 算子，且后两个问题存在已知显式解：

$$\arg\min_{E\in\mathbb{R}^{T\times P}}\left\{\mu\lambda\|E\|_1+\frac{L}{2}\left\|E-G_k^E\right\|_F^2\right\}=\boldsymbol{S}_{\frac{\mu\lambda}{L}}\left[G_k^E\right], \tag{19}$$

$$\arg\min_{N\in\mathbb{R}^{T\times P}}\left\{\mu I_{\text{Box}(\delta)}(N)+\frac{L}{2}\left\|N-G_k^N\right\|_F^2\right\}=\boldsymbol{P}_{\text{Box}(\delta)}\left[G_k^N\right]. \tag{20}$$

其中，$\boldsymbol{S}_\varepsilon[\cdot]$ 表示阈值为 $\varepsilon>0$ 的软阈值算子，$\boldsymbol{P}_M[\cdot]$ 表示到凸集 $M$ 的投影算子。故只需讨论第一个子问题：

$$\min_{A\in\mathbb{R}^{T\times P}}\left\{I_{V_q^{(P)}}(A)+\mu\|A\|_*+\frac{L}{2}\left\|A-G_k^A\right\|_F^2\right\}. \tag{21}$$

本文命题 1 证明该问题亦存在显式解。

**命题 1**：设 $\hat{G}_k^A = \boldsymbol{P}_{V_q^{(P)}}[G_k^A]$ 为矩阵 $G_k^A$ 到子空间 $V_q^{(P)}$ 的投影，且奇异值分解记为 $\hat{G}_k^A = \hat{U}\hat{\Sigma}\hat{V}^T$，则优化问题(21)的解为 $\hat{U}\boldsymbol{S}_{\frac{\mu}{L}}[\hat{\Sigma}]\hat{V}^T$。

**证明**：(21)式目标函数 $h_0(A) = I_{V_q^{(P)}}(A) + \mu\|A\|_* + \dfrac{L}{2}\|A - G_k^A\|_F^2$ 是一个严格凸函数，故存在唯一的极小值点，因此只需证明该极小值点正是

$$A^* = \hat{U}\boldsymbol{S}_{\frac{\mu}{L}}\left[\hat{\Sigma}\right]\hat{V}^T. \tag{22}$$

由于 $\hat{G}_k^A = \boldsymbol{P}_{V_q^{(P)}}[G_k^A] \in V_q^{(P)}$，$\hat{G}_k^A$ 中每个列向量都包含于子空间 $V_J$。由SVD分解关系式 $\hat{U}\hat{\Sigma}\hat{V}^T = \hat{G}_k^A$，得到 $\hat{U} = \hat{G}_k^A\hat{V}\hat{\Sigma}^{-1}$，因此 $\bar{U}$ 中每个列向量是 $\hat{G}_k^A$ 中列向量的线性组合，故包含于 $V_J$。同理，根据(22)式定义，$A^*$ 中每个列向量也包含于 $V_J$，所以 $A^* \in V_q^{(P)}$ 且

$$h_0(A^*) = \mu\|A\|_* + \frac{L}{2}\|A - G_k^A\|_F^2 < +\infty.$$

设 $V_q^{(P)}$ 的正交补空间为 $C_q^{(P)}$，即 $V_q^{(P)} \oplus C_q^{(P)} = \mathbb{R}^{T\times P}$（"$\oplus$"表示线性空间直和）。因此，$G_k^A - \hat{G}_k^A \in C_J^{(P)}$。对任意矩阵 $B \in V_q^{(P)}$，均有

$$\begin{aligned}
h_0(B) &= \mu\|B\|_* + \frac{L}{2}\|(B - \hat{G}_k^A) - (G_k^A - \hat{G}_k^A)\|_F^2 \\
&= \mu\|B\|_* + \frac{L}{2}\|(B - \hat{G}_k^A)\| + \frac{L}{2}\|(G_k^A - \hat{G}_k^A)\|_F^2.
\end{aligned} \tag{23}$$

根据文献[34]中定理 2.1，$A^*$ 为以下优化问题：

$$\min_{A \in \mathbb{R}^{T\times P}} \mu\|A\|_* + \frac{L}{2}\|A - \hat{G}_k^A\|_F^2 \tag{24}$$

的唯一解。故可以得到

$$\begin{aligned}
h_0(B) &\ge \mu\|A^*\|_* + \frac{L}{2}\|(A^* - \hat{G}_k^A)\| + \frac{L}{2}\|(G_k^A - \hat{G}_k^A)\|_F^2 \\
&= \mu\|A^*\|_* + \frac{L}{2}\|(A^* - \hat{G}_k^A) - (G_k^A - \hat{G}_k^A)\|_F^2 \\
&= h_0(A),
\end{aligned} \tag{25}$$

且(25)式取等号当且仅当 $B = A^*$（注意，(23)，(25)式的等价变换只对 $V_q^{(P)}$ 中元素成立）。证毕。

根据以上讨论，仿照 SPCP 问题的 APG 算法，设计 SPCP-MRC 优化问题的 APG 算法：

**算法 1．SPCP-MRC 优化问题的 APG 算法**

**输入**：流量矩阵数据 $X \in \mathbb{R}^{T\times P}$；

**初始化**：$A_{-1} = A_0 = 0^{T\times P}$；$E_{-1} = E_0 = 0^{T\times P}$；$N_{-1} = N_0 = 0^{T\times P}$；$t_{-1} = t_0 = 1$；$k = 0$；

　　　　　$\mu_0 = 0.99\|X\|_2$；$\bar{\mu} = 10^{-5}\mu_0$；$\eta = 0.9$；$L = 3$；$\lambda = 1/\sqrt{\max(T, P)}$.

**While** not converged **do**

$$Y_k^A = A_k + \frac{t_{k-1}-1}{t_k}(A_k - A_{k-1}); Y_k^E = E_k + \frac{t_{k-1}-1}{t_k}(E_k - E_{k-1}); Y_k^N = N_k + \frac{t_{k-1}-1}{t_k}(N_k - N_{k-1});$$

$$G_k^A = Y_k^A - \frac{L}{2}(Y_k^A + Y_k^E + Y_k^N - N); G_k^E = Y_k^E - \frac{L}{2}(Y_k^A + Y_k^E + Y_k^N - N); G_k^N = Y_k^N - \frac{L}{2}(Y_k^A + Y_k^E + Y_k^N - N);$$

$\hat{G}_k^A = \boldsymbol{P}_{V_q^{(P)}}[G_k^A];$  //  $\boldsymbol{P}_{V_q^{(P)}}[\cdot]$ 表示到子空间 $V_q^{(P)}$ 的投影算子

$(\hat{U}, \hat{\Sigma}, \hat{V}) = svd[\hat{G}_k^A];$  //  $svd[\cdot]$ 表示矩阵 SVD 分解

$A_{k+1} = \hat{U} \boldsymbol{S}_{\frac{\mu}{L}}[\hat{\Sigma}] \hat{V}^T;$  //  $\boldsymbol{S}_\varepsilon[\cdot]$ 表示阈值为 $\varepsilon > 0$ 的软阈值算子

$E_{k+1} = \boldsymbol{S}_{\frac{\lambda\mu}{L}}[G_k^E];$

$N_{k+1} = \boldsymbol{P}_{\text{Box}(\delta)}\left[G_k^N\right];$  //  $\boldsymbol{P}_{\text{Box}(\delta)}[\cdot]$ 表示到凸集 Box($\delta$) 的投影算子

$t_{k+1} = (1 + \sqrt{4t_k^2 + 1})/2;$

$\mu_{k+1} = \max(\eta\mu_k, \bar{\mu});$

$k = k+1;$

**End While**

**输出**：分解流量子矩阵 $A = A_k$；$E = E_k$；$N = N_k$.

设 $k$ 表示算法迭代步数，定理 1 证明此 APG 算法具有 $O(1/k^2)$ 收敛速度，其证明思想与文献[35]中定理 4.4 很相似，故本文省略证明过程。

**定理 1**：利用算法 1 求解优化问题(15)，则对目标函数 $\mu g(A,E,N) + f(A,E,N)$ 中松弛参数 $\mu$ 的任意取值 $\bar{\mu} > 0$，当迭代步数 $k > k_0 = \lceil \log(\mu_0/\bar{\mu})/\log(1/\eta) \rceil$ 时，满足

$$F(X_k) - F(X^*) \le 6 \| X_{k_0} - X^* \|_F^2 /(k - k_0 + 1)^2 \tag{26}$$

其中 $X_k = (A_k, E_k, N_k)$ 表示该算法第 $k$ 步迭代的中间结果，$X^* = (A^*, E^*, N^*)$ 表示问题(15)松弛参数取值为 $\mu = \bar{\mu}$ 时的最优解。

# 6  实验与评价

通过模拟实验比较SPCP-MRC与其他几种流量矩阵分解方法。其中，6.1节介绍流量矩阵模拟方法，6.2 节通过定量与定性手段，比较各种流量分解方法的计算结果。

## 6.1  模拟流量矩阵生成

已有研究中使用真实流量矩阵或模拟流量矩阵进行评价。使用真实流量矩阵具有很高说服力，但存在两方面实际困难：一方面，公开的真实流量矩阵数据集比较少，且难以覆盖各种情况；另一方面，真实流量矩阵数据缺乏各流量成分的 Ground-Truth 信息，不利于定量评价。采用模拟流量矩阵的优点包括[36]：（1）模拟实验可以重复多次，有助于发现稳定的实验规律；（2）模拟流量矩阵包含各流量成分 Ground-Truth 信息，易于定量评价；（3）能够灵活控制某些参数，例如异常的类型与幅度、噪声幅度等，评价流量分解方法在不同网络状况下的性能。本文利用丰富的模拟实验，评价和比较多种流量矩阵分解方法。模拟生成的

流量矩阵满足"确定性流量+异常流量+噪声流量"的三元组成假设。噪声流量时间序列服从高斯白噪声，可通过独立的正态分布随机变量来模拟。以下介绍另外两类流量成分的模拟方法。

第一，模拟确定性流量，同时刻画其空间特性与时间特性。对每个 OD 流 $X_p$ $(1 \le p \le P)$，假设确定性流量时间序列 $A_p$ 由非负常数 $a_{p,0}$（表示确定性流量均值）与若干正弦函数组合而成：

$$A_p(t) = a_{p,0} + \sum_{m=1}^{M} a_{p,m} \sin\left(\frac{2\pi l_m}{T} t + \theta_{p,m}\right), \quad 1 \le t \le T. \tag{27}$$

(27)式的组成假设常见于互联网流量建模（如文献[3][9]），以近似流量趋势成分。集合 $\{a_{p,0}\}_{p=1}^{P}$ 刻画确定性流量的空间特性，文献[15]指出近似满足重力模型。本文每次模拟均依重力模型，独立生成一组 $\{a_{p,0}\}_{p=1}^{P}$，且总和等于常量 $10^6$ MB。正弦函数刻画确定性流量的周期性模式属于时间特性，相关参数包括频率参数 $\{l_m\}$、幅度参数 $\{a_{m,p}\}$ 与相位参数 $\{\theta_{m,p}\}$。本文模拟一周内的流量矩阵，最小时间周期为 $\Delta t = 5$ 分钟，则时间周期数目 $T = 7 \times 24 \times 60 / \Delta t = 2016$，且第 $m$ 个 $(1 \le m \le M)$ 正弦函数的周期为 $T \cdot \Delta t / 60 l_m = 168 / l_m$ 小时。文献[5][9]指出正常流量以 24、12、6、3、1.5 小时的周期性模式最明显，故对应频率参数 $\{l_m\} = \{7,14,28,56,112\}$。由真实数据经验，周期性模式的幅度与确定性流量均值成正比，且周期为 24 小时的"日模式"幅度最大，其他周期性模式的幅度随之递减，故本文设幅度参数 $\{a_{m,p}\}$ 满足 $a_{m,p} = 0.5 a_{m-1,p}$ $(1 \le m \le 5)$。此外，假设相位参数 $\{\theta_{m,p}\}$ 是独立同分布的随机变量，均为服从 $[-\pi/5, \pi/5]$ 的均匀分布。

第二，基于合理的模型假设，模拟多种异常流量。所谓异常，是指数据中不符合预期行为的模式[37]。网络异常的种类很多，其中一些会导致流量的显著变化，如Alpha、DoS/DDoS、Flash Crowd、Ingress/Egress Shift等，这些异常称为流量异常。每种流量异常都具备独特的时间特性与空间特性[18]。根据文献[17]中方法，基于流量异常真实数据分析结果，对表 1 总结的配置参数赋以合理的取值，模拟几种典型流量异常。

表 1　模拟异常流量的配置参数及取值

| 参数名称 | 受影响源/目的点数 | 持续时间 | 异常流量幅度 | 异常形状函数 |
| --- | --- | --- | --- | --- |
| 可能取值 | (1, 1), (N, 1), (2, 2) | 分钟，小时 | 乘法因子 $\delta$ | 脉冲、三角、阶梯 |

- **Alpha** 是由点对点大文件传输引发的流量异常，源点和目的点都是唯一的，故异常流量只影响一个 OD 流。在多数情况下，Alpha 异常比较短，本文模拟中持续时间设定为 30 分钟。利用乘法因子 $\delta > 0$ 刻画 Alpha 异常流量大小：异常流量为受影响 OD 流确定性流量均值的 $\delta$ 倍。Alpha 异常流量急剧上升至峰值，故选用阶梯函数刻画其形状。

- **DoS/DDoS** 是对固定目标点的洪泛攻击引发的流量异常，其攻击源可能为一个（DoS）或多个（DDoS）。DoS 异常中攻击源唯一，故异常流量只影响一个 OD 流；DDoS 异常中互联网大量主机被攻击者控制作为攻击源，因此异常流量影响多个具有相同目的点的 OD 流。DoS/DDoS 异常从几分钟到若干天不等，但多集中在 5-30 分钟[17]，本文模拟中持续时间设定为 30 分钟。利用乘法因子 $\delta > 0$ 刻画 DoS/DDoS 异常流量大小。DoS/DDoS 异常流量一般逐渐上升至峰值，然后逐渐减为零，故选用对称斜线函数刻画其形状。

- Flash Crowd 是短时间内大量用户对特定 Web 站点访问激增引发的流量异常。这类异常有的能够预测（如体育比赛、软件更新），有的难以预测（如突发新闻事件）。与 DDoS 异常相似，Flash Crowd 异常影响多个具有相同目的点的 OD 流。Flash Crowd 异常可以任意长，但通常比 Alpha 异常久，本文模拟中持续时间设定为 150 分钟。同样利用乘法因子 $\delta > 0$ 刻画异常流量变化。Flash Crowd 异常流量从零迅速上升到峰值之后缓慢下降到零，用非对称斜线函数（上升段斜率较大，下降段斜率较小）刻画其形状。

- Ingress/Egress Shift 是网络管理中更改 BGP 配置，导致部分 IP 地址发生源/目的点移位引发的流量异常，该异常发生前提是用户可通过多个备选的源/目的点进入/离开网络。Ingress/Egress Shift 异常表现为一个 OD 流流量减少，同时这部分减少的流量添加到另一个 OD 流中，因此影响 2 个源点与 2 个目的点。Ingress/Egress Shift 异常保持到下一次更改 BGP 配置，因此持续时间很长，从几个小时到几天不等，本文模拟中设为 10 小时。利用乘法因子 $0 < \delta < 1$，刻画异常流量变化：第一个 OD 流确定性流量的 $\delta$ 倍移入第二个 OD 流中。Ingress/Egress Shift 异常是瞬时发生和持续保持的，因此利用阶梯函数刻画其形状。

此外，实验中还模拟最简单的 Random Point 流量异常，即每个异常随机影响一个 OD 流，持续一个时间周期，异常流量为所在 OD 流确定性流量均值的 $\delta$ 倍。表 2 总结了各种异常流量的参数选择。

**表 2  各种异常流量配置参数的选择**

| 参数名称 | 受影响源/目的点数 | 持续时间 | 异常流量幅度 | 异常形状函数 |
|---|---|---|---|---|
| Random Point | (1, 1) | 5 分钟 | 乘法因子 | 脉冲 |
| Alpha | (1, 1) | 30 分钟 | 乘法因子 | 阶梯 |
| DoS | (1, 1) | 30 分钟 | 乘法因子 | 对称斜线 |
| DDoS | ($N$, 1) | 30 分钟 | 乘法因子 | 对称斜线 |
| Flash Crowd | ($N$, 1) | 150 分钟 | 乘法因子 | 非对称斜线 |
| Ingress/Egress Shift | (2, 2) | 10 小时 | 乘法因子 | 阶梯 |

将确定性流量矩阵和不同种类的异常流量矩阵以及不同幅度的噪声流量矩阵 ($\alpha = 0.05, 0.1$) 组合，模拟多组流量矩阵。全部模拟流量矩阵如表 3 所示，其中每组流量矩阵由 4 个参数确定，独立生成 50 个样本。

**表 3  实验使用的模拟流量矩阵总结**

| 流量矩阵名称 | 异常参数 | | | 噪声参数 |
|---|---|---|---|---|
| | 类型 | 幅度 $\delta$ | 比率($r$)或数量(#) | $\alpha$ |
| $X_{\text{Random}}(\delta, r, \alpha)$ | Random Point | | $r = 0.01$ | |
| $X_{\text{Alpha}}(\delta, \#, \alpha)$ | Alpha | | # = 500 | |
| $X_{\text{DoS}}(\delta, \#, \alpha)$ | DoS | 0.5 | # = 500 | 0.05, 0.1 |
| $X_{\text{DDoS}}(\delta, \#, \alpha)$ | DDoS, $N=5$ | | # = 100 | |
| $X_{\text{Flash}}(\delta, \#, \alpha)$ | Flash Crowd, $N=3$ | | # = 50 | |
| $X_{\text{shift}}(\delta, \#, \alpha)$ | Ingress/Egress Shift | | # = 10 | |

## 6.2  实验结果

利用 SPCP-MRC 与四种对比方法（PCA、PCP、SPCP、SPCP-FDR），分解 6.1 节模拟的各组流量矩阵。对 PCA，总流量分解为确定性流量和残差流量，确定性流量由总流量矩阵前 11 个主成分向量投影估计（因

为Ground-Truth矩阵的秩为 11），残差流量对应异常流量与噪声流量之和；对PCP，总流量分解为确定性流量和异常流量；对SPCP-MRC、SPCP及SPCP-FDR，总流量分解为确定性流量、异常流量和噪声流量。对流量分解得到的各流量成分，首先定量评价其准确性，然后进行定性分析。

### 6.2.1 流量分解的准确性评价

对任意流量矩阵 $X$，设其三个流量子矩阵真实值为 $(A, E, N)$。对于 SPCP-MRC、SPCP 及 SPCP-FDR，$X$ 的分解结果为 $(\hat{A}, \hat{E}, \hat{N})$，由相对误差定义确定性流量、异常流量及噪声流量分解的准确度：

$$\begin{cases} \text{accuracy}(A) = \dfrac{\| A - \hat{A} \|_F}{\| A \|_F} \\[2mm] \text{accuracy}(E) = \dfrac{\| E - \hat{E} \|_F}{\| E \|_F} \\[2mm] \text{accuracy}(N) = \dfrac{\| N - \hat{N} \|_F}{\| N \|_F} \end{cases} \tag{28}$$

对于PCA方法，$X$ 的分解结果为确定性流量矩阵 $\hat{A}$ 和残差流量矩阵 $R$，残差流量中不区分异常流量与噪声流量，确定性流量分解的准确度与(28)式中定义相同，其他两种流量分解的准确度分别定义为残差流量 $R$ 与真实值 $E$ 和 $N$ 的相对误差。对于PCP方法，分解结果为确定性流量矩阵 $\hat{A}$ 和异常流量矩阵 $\hat{E}$，无噪声成分，只能考虑确定性流量和异常流量分解的准确度，采用(28)式定义。表 4-表 6比较了各种流量分解方法对多组模拟流量矩阵（具有不同异常类型和不同噪声幅度）的准确度，为每组 50 个样本结果的平均值。

表 4 确定性流量分解的准确度 accuracy($A$)

| 流量矩阵 | $\alpha = 0.05$ | | | | | $\alpha = 0.1$ | | | | |
|---|---|---|---|---|---|---|---|---|---|---|
| | PCA | PCP | SPCP | SPCP-FDR | SPCP-MRC | PCA | PCP | SPCP | SPCP-FDR | SPCP-MRC |
| $X_{\text{random}}$ | 0.051 | 0.035 | 0.051 | 0.018 | **0.016** | 0.081 | 0.057 | 0.086 | 0.027 | **0.023** |
| $X_{\text{alpha}}$ | 0.054 | 0.035 | 0.050 | 0.018 | **0.016** | 0.083 | 0.057 | 0.084 | 0.026 | **0.022** |
| $X_{\text{DoS}}$ | 0.044 | 0.035 | 0.049 | 0.017 | **0.015** | 0.077 | 0.057 | 0.083 | 0.026 | **0.022** |
| $X_{\text{DDoS}}$ | 0.045 | 0.035 | 0.049 | 0.018 | **0.015** | 0.077 | 0.057 | 0.084 | 0.026 | **0.022** |
| $X_{\text{flash}}$ | 0.049 | 0.035 | 0.049 | 0.018 | **0.015** | 0.079 | 0.057 | 0.083 | 0.026 | **0.022** |
| $X_{\text{shift}}$ | 0.059 | 0.036 | 0.051 | 0.018 | **0.017** | 0.085 | 0.059 | 0.086 | 0.027 | **0.023** |

表 5 异常流量分解的准确度 accuracy($E$)

| 流量矩阵 | $\alpha = 0.05$ | | | | | $\alpha = 0.1$ | | | | |
|---|---|---|---|---|---|---|---|---|---|---|
| | PCA | PCP | SPCP | SPCP-FDR | SPCP-MRC | PCA | PCP | SPCP | SPCP-FDR | SPCP-MRC |
| $X_{\text{random}}$ | 1.001 | 1.127 | 0.510 | 0.675 | **0.372** | 1.466 | 2.148 | 0.885 | 0.937 | **0.692** |
| $X_{\text{alpha}}$ | 0.983 | 1.010 | 0.495 | 0.317 | **0.295** | 1.394 | 1.922 | 0.869 | 0.552 | **0.550** |
| $X_{\text{DoS}}$ | 1.212 | 1.561 | 0.655 | 0.451 | **0.417** | 1.885 | 2.967 | 0.928 | 0.776 | **0.759** |
| $X_{\text{DDoS}}$ | 1.214 | 1.577 | 0.673 | 0.460 | **0.418** | 1.884 | 3.005 | 0.935 | 0.774 | **0.765** |
| $X_{\text{flash}}$ | 1.133 | 1.321 | 0.688 | 0.339 | **0.312** | 1.622 | 2.510 | 0.944 | 0.632 | **0.595** |
| $X_{\text{shift}}$ | 1.559 | 1.290 | 0.478 | 0.200 | **0.161** | 2.603 | 2.445 | 0.755 | 0.432 | **0.421** |

表 6  噪声流量分解的准确度 accuracy($N$)

| 流量矩阵 | $\alpha = 0.05$ | | | | | $\alpha = 0.1$ | | | | |
|---|---|---|---|---|---|---|---|---|---|---|
| | PCA | PCP | SPCP | SPCP-FDR | SPCP-MRC | PCA | PCP | SPCP | SPCP-FDR | SPCP-MRC |
| $X_{random}$ | 1.008 | N/A | 1.218 | 0.679 | **0.452** | 0.865 | N/A | 1.035 | 0.504 | **0.407** |
| $X_{alpha}$ | 1.043 | N/A | 1.212 | 0.416 | **0.415** | 0.874 | N/A | 1.038 | 0.374 | **0.359** |
| $X_{DoS}$ | 0.887 | N/A | 1.168 | 0.399 | **0.392** | 0.833 | N/A | 0.972 | 0.322 | **0.314** |
| $X_{DDoS}$ | 0.888 | N/A | 1.168 | 0.397 | **0.389** | 0.835 | N/A | 0.970 | 0.321 | **0.313** |
| $X_{flash}$ | 0.901 | N/A | 1.207 | 0.363 | **0.358** | 0.840 | N/A | 0.993 | 0.313 | **0.304** |
| $X_{shift}$ | 0.821 | N/A | 1.169 | 0.324 | **0.316** | 0.818 | N/A | 0.991 | 0.282 | **0.280** |

　　根据上述表格中数据，归纳出几点实验结果：（1）针对各组流量矩阵的三种流量成分分解，SPCP-MRC 准确度均优于几种对比方法，且只有 SPCP-FDR 准确度与之接近；（2）相同噪声幅度下，SPCP-MRC 分解各组流量矩阵中确定性流量的准确度很接近，当 $\alpha = 0.05$ 时不超过 0.017，当 $\alpha = 0.1$ 时不超过 0.023；（3）相同噪声幅度下，SPCP-MRC 分解各组流量矩阵中异常流量的准确度存在较大差异，Ingress/Egress Shift 异常的相对误差最小，其次为 Alpha 异常和 Flash Crowd 异常，Random Point 异常与 DoS/DDoS 异常相对误差较大；（4）SPCP-MRC 分解噪声流量的准确度当 $\alpha = 0.05$ 时介于 0.316 到 0.452，当 $\alpha = 0.1$ 时介于 0.280 到 0.407，其中流量矩阵 $X_{shift}$ 噪声流量分解结果最佳，$X_{random}$ 噪声流量分解结果最差。

　　综上所述，SPCP-MRC 分解三种流量成分的准确度均高于对比方法；对相同噪声幅度下含不同种类异常的流量矩阵，SPCP-MRC 分解确定性成分的准确度很接近，分解异常成分与噪声成分的准确度存在差异。

### 6.2.2　流量分解结果的定性分析

　　本文除定量评价准确度以外，还对各种流量分解方法的实验结果进行定性分析。首先，比较确定性流量成分的分解结果。图 1 展示各种流量矩阵分解方法对流量矩阵 $X_{random}$ 中某个OD流的确定性流量时间序列恢复结果（对于其他种类的流量矩阵，分解确定性流量结果很相似，故从略）。由图可见，SPCP-MRC 分解确定性流量时间序列与 Ground-Truth 数据匹配度最高，且连续光滑变化；SPCP-FDR 分解结果也具有比较小的误差，但是存在小幅度波动，不具有光滑性；PCA、PCP、SPCP 分解结果的误差明显，其中 PCA 分解结果包含少数显著偏离 Ground-Truth 数据的样本，可以解释对应时刻总流量中包含异常流量且未被有效分离，SPCP 分解结果对确定性流量波峰、波谷模式的匹配效果不佳。

　　其次，比较多种异常流量成分的分解结果。图 2-图 6 分别展示各种流量矩阵分解方法对流量矩阵 $X_{random}$、$X_{alpha}$、$X_{DDoS}$、$X_{flash}$ 和 $X_{shift}$ 中异常流量时间序列的恢复结果（图中均考虑不同的噪声幅度，左半边为 $\alpha = 0.05$，右半边为 $\alpha = 0.1$）。每个 Ingress/Egress Shift 异常影响一对OD流，故图 6 同时展示流量的增加（上半部分）与减少（下半部分）。对于其他种类异常，只展示一个受影响的OD流（尽管DDoS异常与Flash Crowd异常影响多个OD流，但各OD流中异常分解结果相似故仅展示一个；DoS异常是DDoS异常的特殊情况，故实验结果从略）。由图可见，SPCP-MRC对各种异常流量成分恢复效果都较为合理，并且在无异常位置出现误判的情况较少；PCA、PCP尽管对真正异常恢复比较准确，但误判现象很严重，特别是在噪声幅度比较大时；随着噪声幅度增加，各种流量分解方法恢复异常流量的效果都有下降，SPCP和SPCP-FDR对真实异常

流量产生明显低估，而SPCP-MRC计算结果所受影响较小。在模拟的各种异常流量中，SPCP-MRC恢复Alpha和Ingress/Egress Shift与对应Ground-Truth异常流量匹配最佳，这可以解释为这两种异常均持续较长时间，且异常出现后变化缓慢，因此与噪声具有较大差异，易于被准确分离。

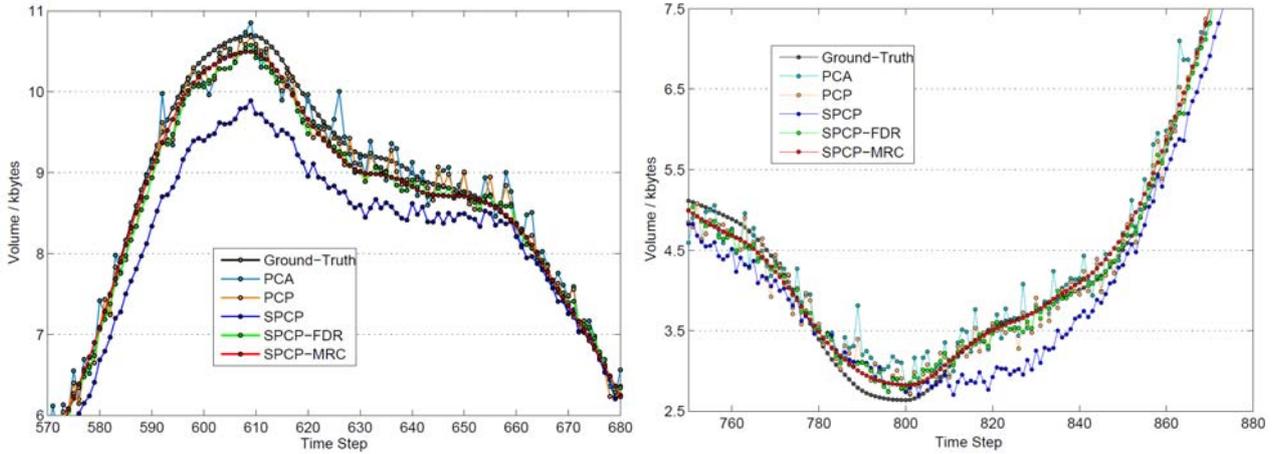

图 1　　各种流量分解方法对确定性流量时间序列的恢复（左为波峰时间段，右为波谷时间段）。

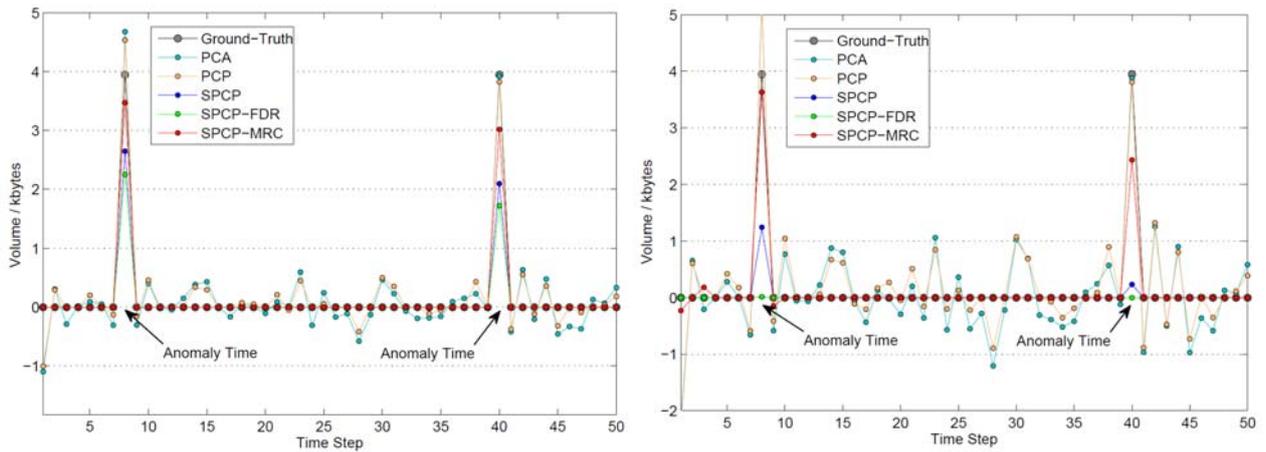

图 2　　各种流量分解方法对 Random Point 异常流量时间序列的恢复（左$\alpha$=0.05，右$\alpha$=0.1）

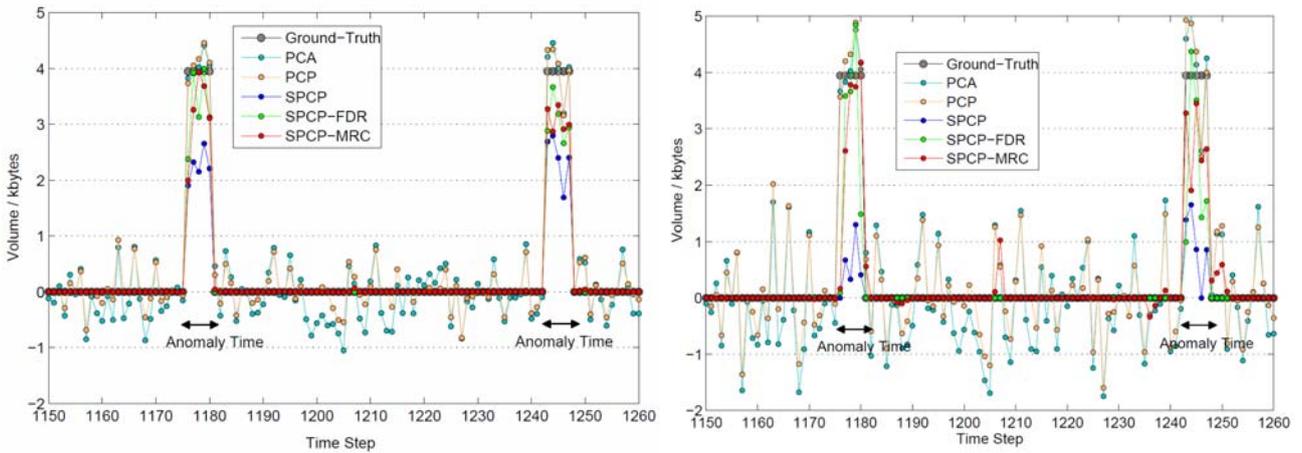

图 3　　各种流量分解方法对 Alpha 异常流量时间序列的恢复（左$\alpha$=0.05，右$\alpha$=0.1）

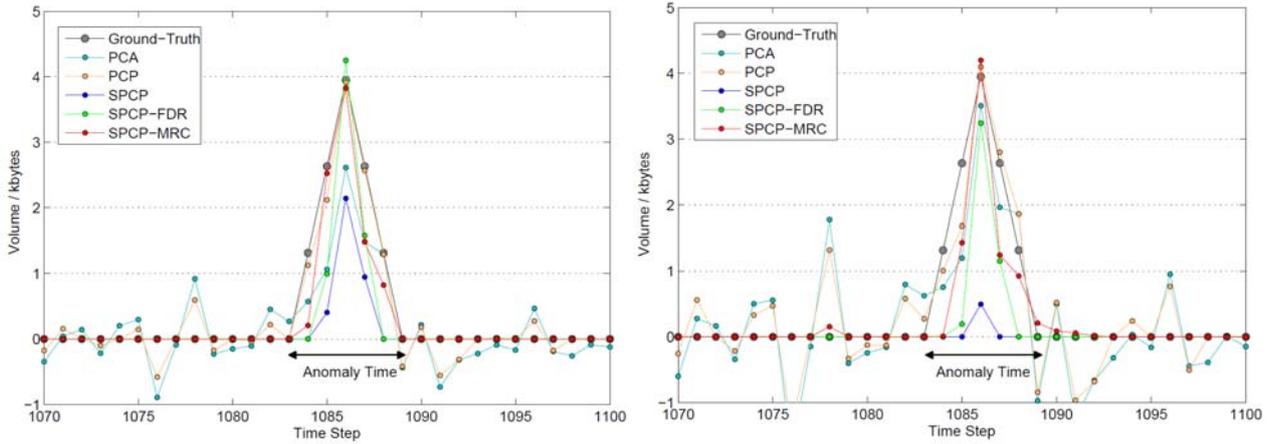

图 4    各种流量分解方法对 DDoS 异常流量时间序列的恢复（左 $\alpha=0.05$，右 $\alpha=0.1$）

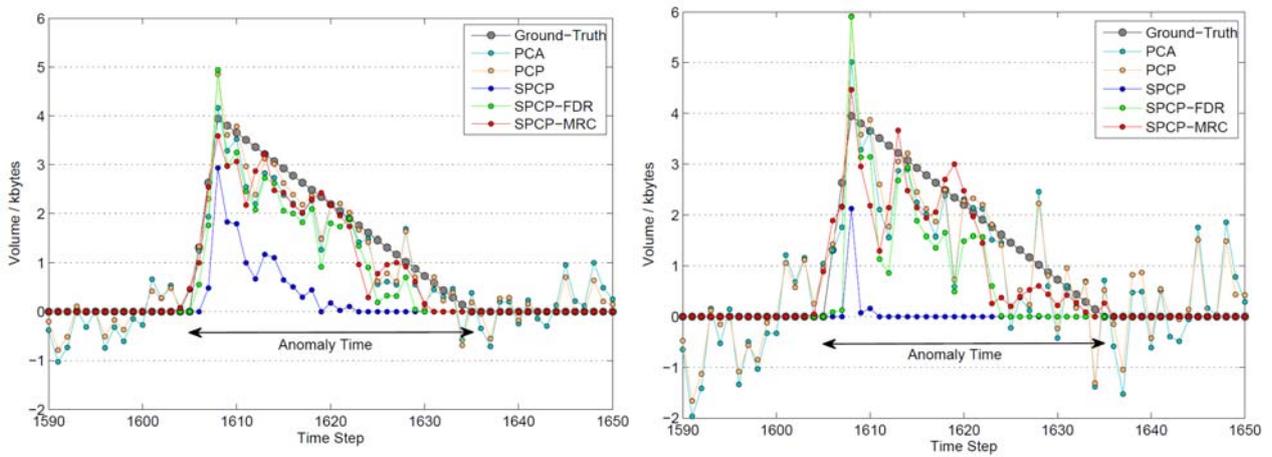

图 5    各种流量分解方法对 Flash Crowd 异常流量时间序列的恢复（左 $\alpha=0.05$，右 $\alpha=0.1$）

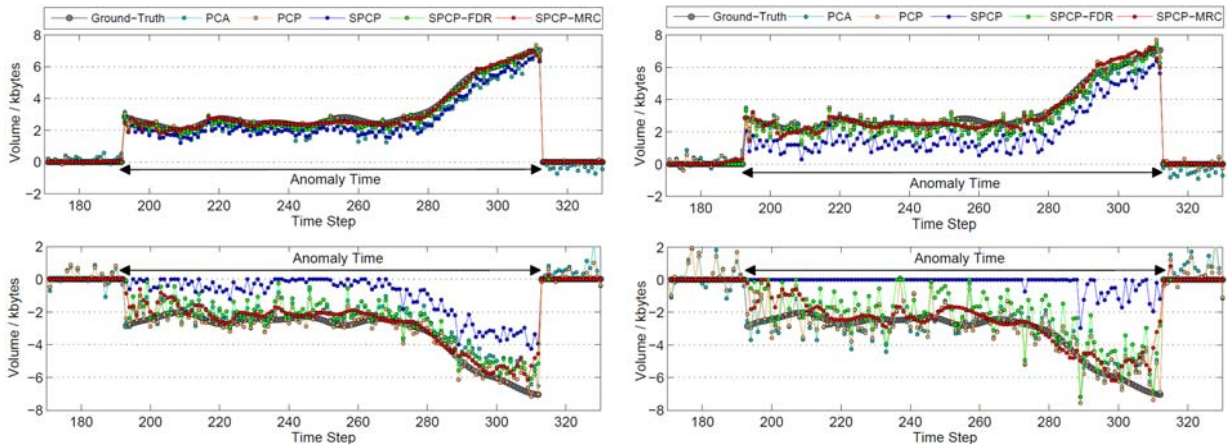

图 6    各种流量分解方法对 Ingress/Egress Shift 异常流量时间序列的恢复（左 $\alpha=0.05$，右 $\alpha=0.1$）

# 7    结论

本文结合 RPCA 理论和多分辨率分析工具，提出基于多分辨率 RPCA 的互联网流量矩阵结构分析框架。建立多分辨率流量矩阵分解模型，提出新的流量矩阵分解优化方法 SPCP-MRC，并为其设计 APG 数值算法，该算法具有 $O(1/k^2)$ 收敛率。针对 APG 算法中确定性流量的优化子问题，本文给出并严格证明了解析

解，该结论是 Proximal 算子研究中的新成果。实验中使用多组模拟流量矩阵，比较多种流量矩阵分解方法。实验结果表明：（1）从定量角度评价，SPCP-MRC 分解三种流量成分的准确度均高于对比方法，且在相同噪声幅度下，SPCP-MRC 分解含不同种类异常流量矩阵，确定性成分的准确度很接近，异常成分与噪声成分的准确度存在较大差异；（2）从定性角度评价，SPCP-MRC 分解结果也更具合理性，表现为对确定性流量的光滑匹配和对异常流量的准确恢复且较少误判。本研究为分析流量矩阵这种重要的互联网高维数据，提出了行之有效的理论与技术手段。

## 参 考 文 献

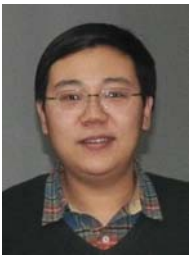

**WANG Zhe**, born in 1984, PhD. His research interest includes computer networks and machine learning.

**HU Kai**, born in 1963, PhD, Associate Professor. His research interest includes embedded systems, parallel and distributed computing.

**YIN Baolin**, born in 1951, PhD, Professor. His research interest includes distributed applications.

## Background

The Internet traffic data analysis plays a significant roll in network management. Recently, the developments of Internet measurement technology enable the network manager to collect the network-wide traffic data more efficiently. As a kind of network-wide traffic data, the traffic matrix has been widely studied for many network management tasks, such as traffic engineering, capacity planning, and anomaly detection. Compared with single-link traffic data, it contains more comprehensive behavior information of the whole network; therefore, analyzing traffic matrix data could help to achieve a global optimal network resource allocation, as well as to track the network-wide anomalies.

Generally, the Internet traffic is considered as the mixture of several unknown traffic components, which correspond to different network behaviors. However, for a specific network management task, usually only one traffic component is the main issue. For example, in capacity planning the deterministic traffic is the most important; while in anomaly detection the anomaly traffic is the primary input. Consequently, estimating these traffic components i.e. the structural analysis of traffic matrix is a central problem in traffic matrix analysis, and an efficient decomposition scheme could provide material support to network management.

In existing works on traffic matrix decomposition, the Principal Component Analysis (PCA) method is widely studied. There exit two limitations of the PCA-based traffic decomposition: 1) the estimation error substantially increases when the network contains large anomalies; 2) this method does not utilize the temporal properties of the traffic data. In this study, we try to fill these two gaps by using the Robust Principal Component Analysis (RPCA) theory and the Multi-Resolution Analysis (MRA) tool, respectively.

This work is supported by the National Science Foundation of China (Nos. 61073013 and 90818024) and the Open Project of State Key Laboratory of Software Development Environment (Nos. SKLSDE-2012ZX-15).